\documentclass[reprint,aps,prx,superscriptaddress,longbibliography,floatfix,nofootinbib]{revtex4-2}

\usepackage{xcolor} 
\newcommand{\rev}[1]{\textcolor{black}{#1}}
\usepackage{graphicx} 
\usepackage{amsmath}
\usepackage{amssymb}

\begin{document}

\title{Template-free search for gravitational wave events using coincident anomaly detection}
\author{Daniel Ratner}
\affiliation{SLAC, Menlo Park, CA, 94110}
\date{\today}


\begin{abstract}
Gravitational-wave (GW) observatories have used template-based search to detect hundreds of compact binary coalescences (CBCs). However, template-based search cannot detect astrophysical sources that lack accurate, computationally-tractable waveform models. Here, we present a novel approach for template-\textit{free} search using coincident anomaly detection (CoAD). CoAD requires neither labeled training examples nor background-only training sets, instead exploiting the coincidence of events across spatially separated detectors as the training loss itself: two neural networks independently analyze data from each detector and are trained to maximize coincident predictions. 
Additionally, we show that integrated gradient analysis can localize GW signals from the neural-network weights, providing a path toward data-driven template construction of unmodeled sources, and further improving precision by frequency matching. Using the CodaBench dataset of real LIGO backgrounds with injected simulated CBCs and sine-Gaussian low-frequency bursts, CoAD achieves recall up to 0.91 and 0.85 respectively at a false-alarm rate of one event per year, and achieves recall above 0.5 at signal-to-noise ratios below 10.  The fully-unsupervised nature of CoAD makes it especially well-suited for next-generation detectors with greater sensitivity and associated increases in GW event rates.
\end{abstract}

\maketitle


\section{Introduction}

The LIGO gravitational wave observatory comprises two interferometers, one in Hanford, Washington, and one in Livingston, Louisiana. The geographic separation of 3000~km serves two purposes: triangulating source position and, critically for this work, distinguishing genuine gravitational wave (GW) signals from local noise. Seismic vibrations, scattered light, thermal fluctuations, and instrumental artifacts are all uncorrelated between sites. Only true astrophysical GWs, propagating at the speed of light, will be synchronized within a 10 millisecond window at both detectors. Standard template-based search exploits this coincidence by searching for waveforms at both sites, and have detected hundreds of Compact Binary Coalescences (CBCs) to date \cite{abac2025gwtc}. Machine-learning (ML) methods have also been applied to GW detection, primarily to reduce computational cost \cite{baker2015multivariate, george2018deep, George2018PLB, marx2025machine}. \rev{However, both template-based search and supervised ML methods require accurate GW models computed from general relativity. Sources for which such models are unavailable or computationally intractable --- e.g. core-collapse supernovae~\cite{dimmelmeier2008gravitational} or even entirely unexpected phenomena --- require template-\emph{free} detection.} 
To date no such unmodeled event types have been found.


Here, we introduce a template-free approach that inverts the traditional paradigm: instead of using coincidence as either a model input or a post-processing filter, we use it as the loss function itself. Coincident anomaly detection (CoAD)~\cite{humble2024coincident} trains two neural networks --- one per detector --- by maximizing agreement between the predictions. Because detector noise is statistically independent \rev{\cite{abbott2016characterization}}, only a data partition that separates correlated GW signals from uncorrelated backgrounds will produce better-than-random prediction agreement. The paper is arranged as follows: a brief survey of existing deep-learning approaches to \rev{template and template-free} search; a description of the CoAD method as applied to GW detection; a case study using the Codabench challenge dataset, including exploration of interpretability of the deep-learning method.

\section{Related Work}

Template-free GW detection methods range from classical signal processing to modern deep learning. Classical methods include the coherent WaveBurst (cWB) algorithm~\cite{klimenko2008coherent}, which identifies excess power in time-frequency representations, and BayesWave~\cite{cornish2021bayeswave}, which models signals as sums of wavelets within a Bayesian framework. Other methods use principle component analysis \cite{powell2015classification, powell2017classification} and k-nearest neighbors \cite{benkHo2020find}.  Deep learning approaches fall into two categories: \textit{Supervised methods} train on simulated GW events to learn specific event types. MLy~\cite{skliris2024toward} trains neural networks on diverse GW morphologies, \cite{mcginn2021generalised} using a generative adversarial network to generalize training data, and \cite{mukund2017transient} uses a partially-supervised boosted neural network trained on wavelets. Supervised methods have high learning potential, but require assumptions about signal characteristics (duration, frequency range, morphology) and may miss phenomena absent from the training distribution. \textit{Anomaly detection methods} instead frame detection as unsupervised outlier identification. Autoencoder networks trained on background noise learn to compress and reconstruct normal detector behavior, and when presented with novel GW signals produce high reconstruction error~\cite{morawski2021anomaly, moreno2022source, raikman2024gwak}. Autoencoders have two potential limitations: First, they require background-only training sets, which may be problematic for future observatories that are sensitive to large numbers of events. Second, learning full noise distributions is harder than the supervised approach of directly learning decision boundaries.

CoAD addresses these limitations by using coincidence itself as the training mechanism. CoAD maintains the strong learning capabilities of supervised methods, while maintaining the model-independence of anomaly-detection methods. Unlike autoencoders, CoAD requires no background-only training set, and unlike supervised methods, it makes no assumptions about signal characteristics. While CoAD contains some similarities to MLy \cite{skliris2024toward} in the use of correlation, the two methods differ fundamentally in the usage: whereas MLy provides data from both detectors (including their correlation) as \textit{input} to the models, in CoAD the correlation is used only within the loss function, and each of the two models receives input data from a single detector.  Consequently, in CoAD it is possible to make predictions when only a single detector is available, as well as to use correlations for post-analysis. We emphasize that CoAD can be complementary to existing methods, especially for high-event-rate scenarios where background-only datasets are unavailable, and for detecting populations of low signal-to-noise events. Given the difference in approach, CoAD can learn to detect different types of anomalies compared to existing methods, making CoAD a potentially useful addition to the existing quiver of methods.

\section{Coincident anomaly detection}

CoAD is designed for tasks that satisfy one key condition: features can be split into two datastreams which are uncorrelated under normal conditions, but which both contain signatures of anomalous events. Each datastream is fed to a model --- typically but not necessarily a neural network (NN) --- which outputs the probability that an event is anomalous.  CoAD defines a loss function, $\hat{F}_\beta$, which rewards the models for making consistent predictions; crucially, $\hat{F}_\beta$ only rewards models with better-than-random agreement, separating correlated anomalies from uncorrelated backgrounds; accidental coincidence of false positives is not rewarded so long as the backgrounds are uncorrelated and occur at the level expected under the assumption of independence. Most importantly, the training is entirely unsupervised --- there is no need even for a training set of normal data.

LIGO's geometry naturally satisfies CoAD's requirements given the two data streams, one from Hanford (H), and the other from Livingston (L). The two detectors, separated by 3000~km, experience statistically independent local noise. GW signals produce the only correlated events which would arrive within a 10 ms window at both sites. Consequently, if the models can learn to identify GW events, they will find a partition which optimizes $\hat{F}_\beta$. Additionally, the model should learn to ignore noise, glitches, and other uncorrelated events which are not rewarded by the $\hat{F}_\beta$ score. Previous studies have considered various possible constructions of $\hat{F}_\beta$ \cite{humble2024coincident, oshea2024coincidence, liang2025coincident}; here, we follow the correlation/covariance structure \cite{oshea2024coincidence}. A schematic of the CoAD concept is shown in Fig.~\ref{fig:schematic}, and details are given in Appendix~\ref{app:loss}.

Once trained, the two models can be applied independently to the H and L datestreams in inference mode. However, to achieve high precision, the two detector outputs can also be combined, i.e.\,using the assumption of coincidence across the two detectors not just for training, but also at inference time. The two predictions can be combined in various ways (see Appendix~\ref{app:inference}); here, we choose a logical AND operation, such that a window is identified as a GW event only if both network outputs are above a chosen threshold. As an unsupervised method, the networks can in principle provide meaningful predictions even on the training data; in addition to the $\hat{F}_\beta$ loss function, CoAD can also provide unsupervised analogs of precision ($\hat{P}$) and recall ($\hat{R}$), which can evaluate performance and check for overfitting on a validation set. (See Appendix~\ref{app:model_selection} for details on these CoAD metrics.) Nonetheless, best practice would be to always run inference on a hold-out test set. We therefore assume a deployed model would use k-fold cross-validation, i.e. repeating the training process k-times on different partitions of the dataset.

\begin{figure}[htbp]
    \centering
    \includegraphics[width=\linewidth]{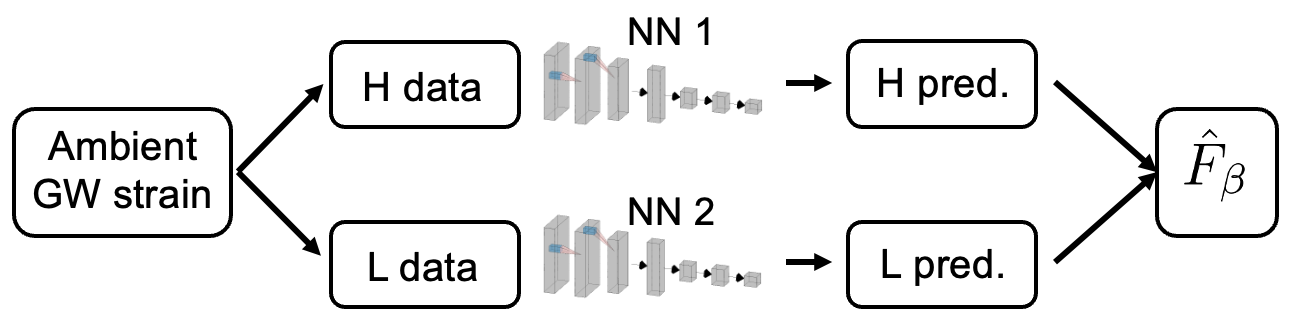}
    \caption{Schematic of the CoAD concept applied to template-free search at LIGO. H and L data are fed into separate NNs, which each make predictions of the probability of an anomaly (H pred.\ and L Pred.\ respectively).  During training, the two predictions are combined in the $\hat{F}_\beta$ loss function, which updates the NNs via back-propagation. At inference time, the networks operate independently, and predictions --- if both are available --- can be combined to improve precision. }
    \label{fig:schematic}
\end{figure}

\section{Results}

We demonstrate the application of CoAD to LIGO by creating a simulated dataset using the LIGO Codabench challenge \cite{campolongo2025building}.  The challenge consists of three datasets, each with 100k examples: one set contains real background data, one contains simulated binary black hole (BBH) events, and one contains simulated sine-Gaussian low frequency (SGLF) events.  The Codabench team constructed the BBH and SGLF data sets by injecting a simulated event onto a real background event. From these three datasets, we create a test challenge by adding a small number of BBH or SGLF events to a large set of backgrounds. The goal for CoAD is to separate the data into normal (i.e. background) and anomalous (BBH or SGLF) events. The public dataset does not include signal-to-noise (SNR) levels for the events, so for the final results presented here we have used a smaller dataset of 10k signals --- paired with SNR levels for each event --- obtained directly from the Codabench authors. \rev{We highlight that the injected simulated events are used solely for the benchmarking demonstration; knowledge of these signal morphologies is not used in training, and a future deployment would train directly on production data with no need for simulations of any kind. In this sense, CoAD is a fully template-free approach.} 

To apply CoAD to our constructed GW challenge, we divide the strain data into two subsets, one containing H data and one containing L data. In principle, the networks can be independent, and indeed could have different architectures (as is the case in some other applications of CoAD \cite{humble2024coincident, oshea2024coincidence, liang2025coincident}). Given the similar noise profiles of the cleaned Codabench data, for simplicity we use identical architectures and weights in this study. 
In principle, CoAD could be applied directly to continuous strain data. However, because CoAD training itself requires observation of anomalous events, the number of training steps needed scales with the observed anomaly rate. Using the Codabench challenge window lengths, and assuming 50\% up-time, there are approximately three billion 50ms windows per decade. A novel GW event class that appears 10 times per year --- approximately one tenth the CBC rate observed in LIGO's 4.0a run \cite{abac2025gwtc} --- would appear at a rate of less than one event per $3 \times 10^{7}$ normal windows, an impractically low anomaly rate. We follow the approach of the original CoAD accelerator task \cite{humble2024coincident, liang2025coincident} and assume a trigger step to remove the vast majority of samples that are likely to be labeled as normal. In the LIGO task, for example, consider a trigger that identifies the 0.4\% most anomalous examples.  Because the two detectors are statistically independent, keeping only windows that pass both L and H triggers would enrich the dataset by a factor of $6 \times 10^4$, resulting in an anomaly rate of approximately 0.2\% for a morphology with 100 events. 

\begin{figure}[htbp]
    \centering
    \includegraphics[width=0.9\linewidth]{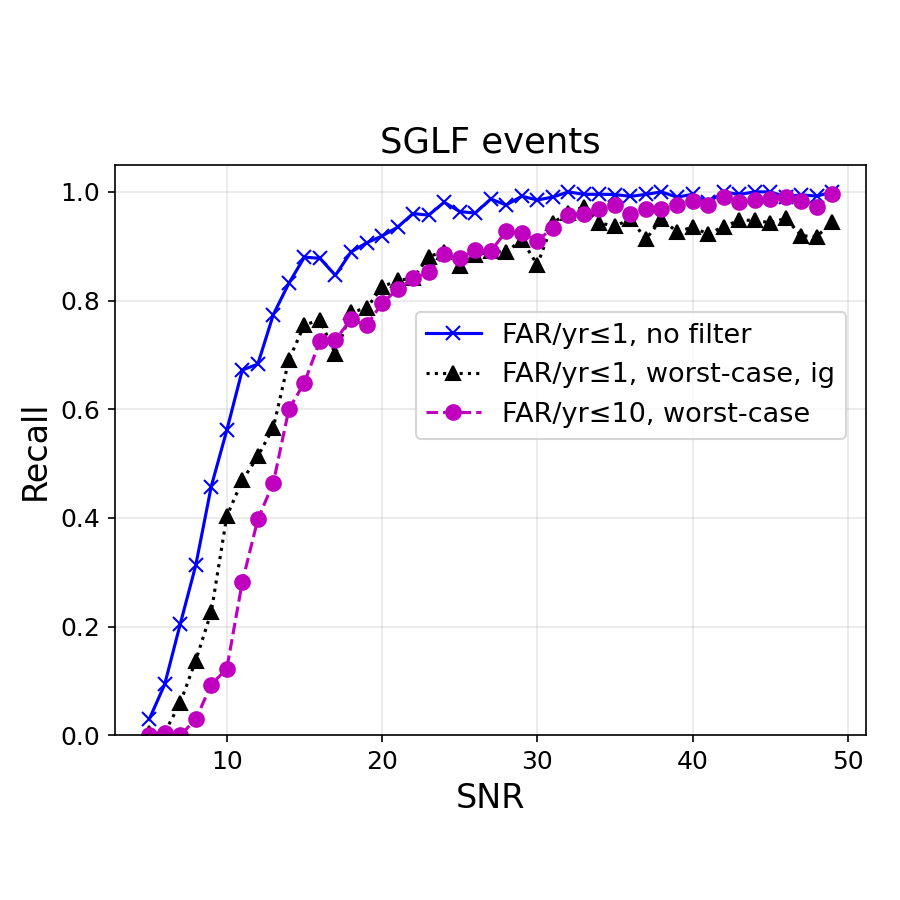}
    \includegraphics[width=0.9\linewidth]{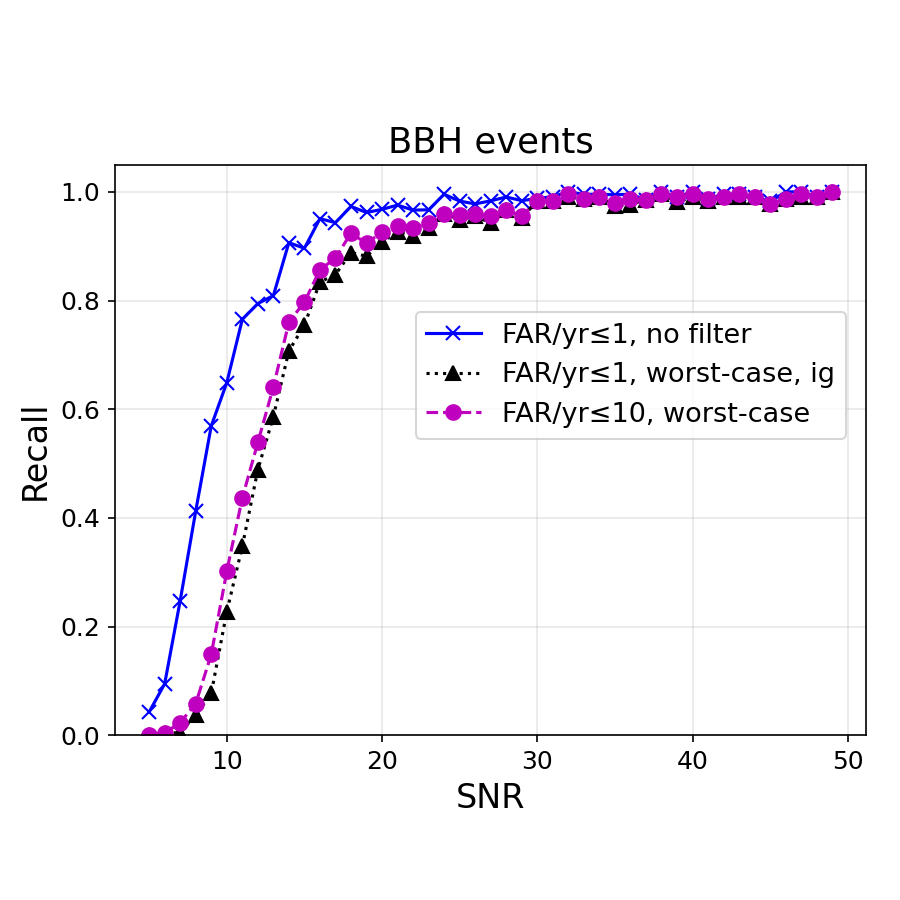}
    \caption{Recall as a function of \rev{test-set} SNR for the 100 event SGLF and BBH models. Blue crosses show recall at FAR-1/year, assuming random backgrounds. Performance drops assuming a worst-case trigger selecting difficult backgrounds, but recovers with either a higher FAR rate (magenta circle) or inclusion of an IG filter (black triangles).}
    \label{fig:recall_vs_snr}
\end{figure}

Following the trigger concept above, we create datasets consisting of 100 SGLF or BBH GW events and 50k backgrounds. We have also experimented with lower anomaly rates; due to the size limitation of the Codabench set, in this case we still reserve 50k background events for training, and resample backgrounds more frequently to match rates as low as 0.02\% anomaly rate (see Appendix \ref{app:dataset}). We transform the cleaned 50ms windows using a short-time Fourier transform.  Sampling from the two sets is stochastic, so some batches contain only background events, depending on the anomaly rate and batch size. The architecture, loss, transform, and training details are given in Appendices~\ref{app:dataset}-\ref{app:loss}. To estimate the recall, we test on the remaining $\sim$10k GW events and 50k background events. Because we only have 50k background events, we approximate the precision by calculating the probability of a single-channel false positive as a function of threshold, and then --- assuming independence of the detectors for background events --- estimate the probability of a simultaneous false positive on both channels. As a metric of model performance, we present the area under the precision-recall curve (PR-AUC) and the recall at a false-alarm rate of 1/year (FAR-1/yr). The PR-AUC is a standard anomaly detection metric that captures performance across a range of precision-recall tradeoffs. Within the GW community, it is common to show FAR-1/yr recall as a function of signal-to-noise ratio (SNR) (e.g. see \cite{raikman2024gwak}). Table \ref{tab:results} shows the recall for various models at a FAR of 1/year, while Fig.~\ref{fig:recall_vs_snr} shows the dependence on SNR. Results are given in Table ~\ref{tab:results}. Notably, CoAD achieves approximately the same performance as direct supervision (i.e. training with cross-entropy) on the same training set, highlighting the strong learning capability of CoAD.   

We show results for both BBH and SGLF events, but for ablation studies we focus on the more challenging SGLF dataset. First, because the chosen anomaly detection rate is somewhat arbitrary, we have explored performance as the rate decreases.  We have found it is possible to achieve similar performance with an anomaly rate as low $2 \times 10^{-4}$, albeit with increased computational expense. We have not explored lower rates, but expect training is possible with sufficient computational resources. 

\rev{Finally we highlight a limitation of the dataset: The Codabench challenge includes only static noise, while realistic data also contains non-Gaussian transient noise (glitches). Though CoAD should be capable of identifying glitches as non-signal (due to lack of correlation between detectors), this capability has yet to be demonstrated. Future CoAD studies should include glitches in the training data.}


\begin{table}
    \centering
    \begin{tabular}{|c|c|c|c|c|c|c|}
    \hline
     \multicolumn{3}{|c|}{Training dataset} & \multicolumn{2}{c|}{PR-AUC} & \multicolumn{2}{c|}{Recall, FA/yr $\leq$ 1} \\
    \hline
     GW & \# of & Event & No & WC trigger & No & WC trigger \\[-3pt]
     type & events & rate & trigger & + IG filter & trigger & + IG filter \\
    \hline
       BBH & 100  & 0.2\% & 0.91 & 0.80 & 0.88 & 0.81 \\
       SGLF & 100 & 0.2\%   & 0.89  & 0.81 
       & 0.86 & 0.81 \\ 
       SGLF & 100 & 0.02\%  & 0.83  & 0.79 
       & 0.80 & 0.77 \\
       SGLF & 50 & 0.2\% & 0.54 & 0.42 
       & 0.53 & 0.42 \\ 
       SGLF & 100 & CE  & 0.86  & 0.73 
       & 0.83 & 0.68 \\
       \hline
    \end{tabular}
    \caption{Summary of performance for the three models presented, including the precision-recall area under the curve (PR-AUC) and the recall at a false-alarm rate below 1/year (`FA/yr $\leq$ 1'). Both statistics are presented for the case of no trigger, and for a worst-case trigger (perfectly correlated to the model) combined with post-filtering with integrated-gradients.  The final row shows the performance of a supervised model trained using cross-entropy (CE) loss, showing similar performance to the CoAD models.
    }
    \label{tab:results}
\end{table}


\begin{figure}[htbp]
    \centering
    \includegraphics[width=.9\linewidth, trim=0 5 0 0]{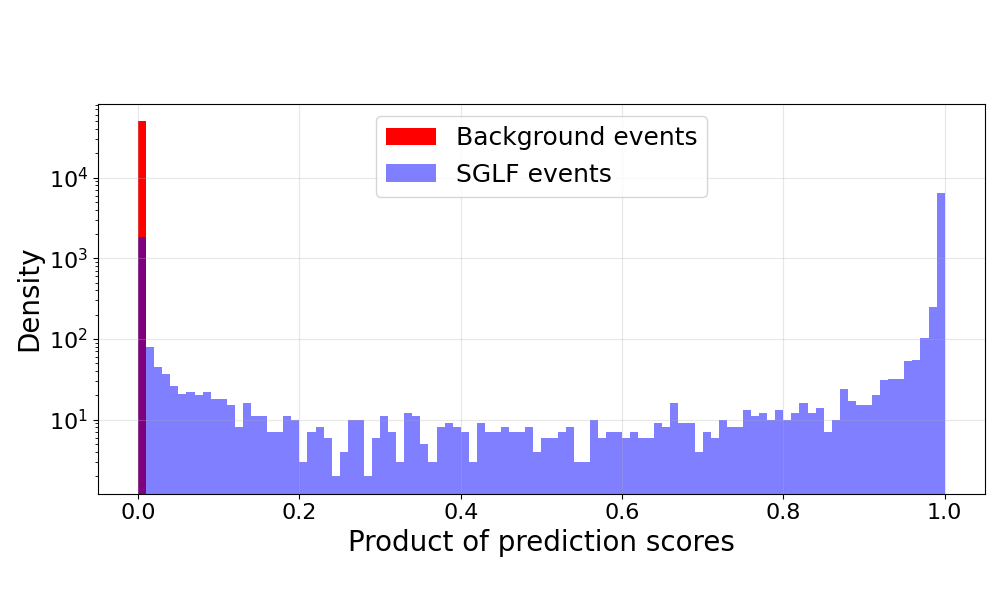}
    \includegraphics[width=.9\linewidth, trim=0 0 0 25]{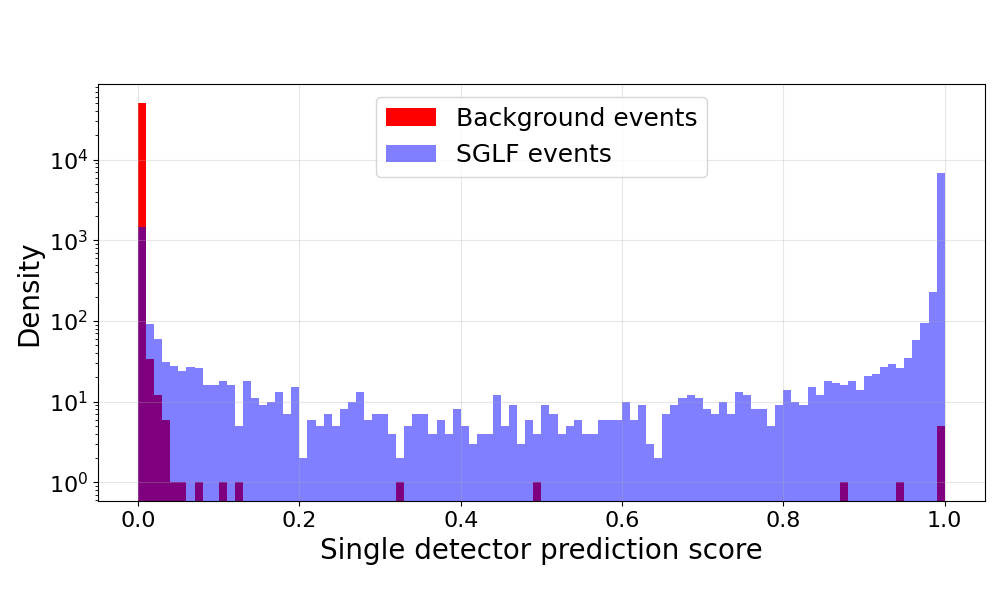}
    \caption{Distribution of predictions for the 0.2\%, 100 event model for a single model output (top) and the product of both models' outputs (bottom), highlighting how combining the two predictions improves precision. Both outputs are passed through a sigmoid function. Note that results presented use a logical AND operation between the two outputs, rather than the product of the sigmoids shown at bottom for visualization purposes.}
    \label{fig:hist}
\end{figure}


\section{Interpretability}

\begin{figure}[htbp]
    \centering
    \includegraphics[width=0.364\linewidth]{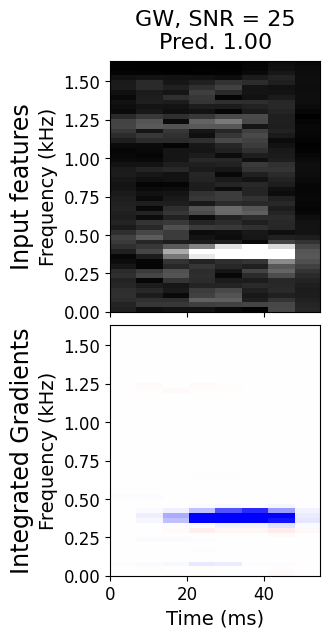}
    \includegraphics[width=0.26\linewidth]{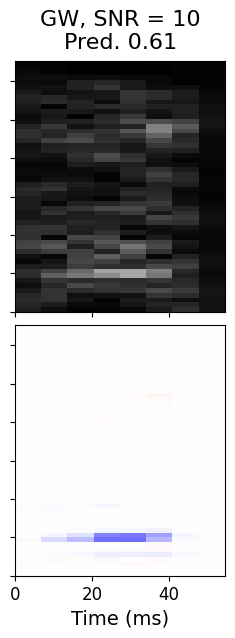}
    \includegraphics[width=0.34\linewidth]{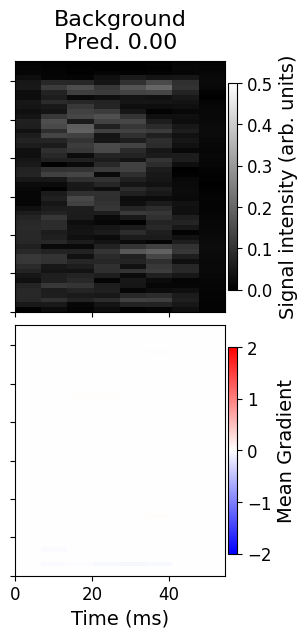}       \caption{IG visualization for three input examples: a strong SGLF at left, a weak SGLF event in the center, and a background event at right. Note that nearly the entirety of the AUC curve has a threshold $\ll 0.1$, so the middle prediction represents a less confident --- but correct --- prediction that the example has a SGLF event.}
    \label{fig:ig}
\end{figure}

A purely black-box identification of an unmodeled event would ultimately be unsatisfying if the nature of the event itself is not revealed.  Ideally, after CoAD identifies a GW partition, researchers could identify the source and produce a physical template from the AI-selected events. However, if the events are near the signal-to-noise limit, the examples may not be readily explainable by eye, and furthermore may be mixed with false positive backgrounds, complicating human analysis. \rev{One solution is to interrogate the NN decisions to extract physically interpretable time-frequency features associated with candidate signals.} For example, various interpretability methods \cite{antwarg2021explaining, borji2019saliency, selvaraju2016grad} can provide insight into how the inputs to a NN affect the output decisions; features with high scores indicate the important components of the raw signal. As a test, we have implemented the integrated gradients (IG) method \cite{sundararajan2017axiomatic}.  IG is similar to saliency \cite{borji2019saliency}, which calculates the gradient of the output relative to the input features, but IG improves stability by integrating the gradients as the input varies from a standard template (in our case a uniform input) to the true input. Fig.~\ref{fig:ig} shows the results for a few examples, with IG identifying a reasonable GW signal region in the time-frequency representation.  \rev{We also note that while this paper uses time-frequency input features, a different choice of inputs could result in a standard time-domain template.}

The IG maps can also serve a second purpose of increasing the precision of the predictions. In standard template matching, the signals must be coincident not just in time, but also in the frequency domain and phase. While the time-windows implicitly enforce the time-coincidence, the IG output can be used analogously to check and enforce frequency coincidence. For example, consider the results of Fig.~\ref{fig:ig}: we can identify the signal's frequency range for each detector (centered at 0.4 kHz for the left plot and 0.25 kHz for the middle plot). We could therefore rule out this window as a coincident signal --- according to the NN's logic --- due to the mismatch in frequencies of the IG map. 

\section{Impact of trigger}

Finally, we consider the impact on performance after including the trigger behavior. The results described above assume a best-case scenario using a random selection of 100k backgrounds. Conversely, the worst-case performance occurs if the trigger is biased to pass those background events which are most likely to deceive the model; the more challenging post-trigger backgrounds could result in more false positives, and thus a lower precision. With only 50k background test events, it is not possible to test this directly: the desired trigger rate would only leave a single background event for training. However, we can exploit the independence of background events to test the implications of triggering on the results of Table~\ref{tab:results}. We explored the worst-case scenario by selecting the 0.4\% of background events predicted to have the highest GW probability by the trained network. (With 50k backgrounds in the test set, this corresponds to 200 events for each detector.) We then estimate the probability of simultaneous false positives, assuming independence of the detectors, based on the single-channel false-positive rate. The impact to performance is highly dependent on the model, with wide variability within different models trained on the same dataset.  However, in general the performance recovers to close to the original level by including the IG post-filter requiring the peak signal to be within 0.05 kHz. Results of the trigger and IG filter are given in Fig.~\ref{fig:recall_vs_snr} and Table~\ref{tab:results}.

While the low FAR is critical if the alarms will be used to trigger multi-messenger observations, CoAD may be additionally useful as a low-SNR discovery method: i.e. CoAD could identify a set events with high confidence when assessed that the set contains true events, but with low confidence for any individual event.  For example, consider a scenario in which CoAD identifies 100 potential events from a set of 25k examples, with a predicted false positive rate of 0.2\% (i.e. 50 expected false positives) based on the number of \rev{independent} single channel positives.  While any individual event is equally as likely to be a false positive as a GW event, collectively such an observation of 100 coincident predictions would be a $>$5-sigma event. Figure~\ref{fig:recall_vs_snr} shows performance for looser bounds (FAR $\leq$ 10/year) under a worst-case trigger.

We have also implemented a simplistic trigger based on the maximum value of the STFT input to evaluate the feasibility of the trigger concept. 
When targeting 0.4\% background retention, the simplistic max-value trigger retains $77$\% of signal events. Moreover, the triggering is only needed for training, and once trained, the neural network would take only 5 hours on a single T4 GPU to process the entire (triggerless) dataset of three billion windows in inference mode. On the other hand, higher retention would result in more events available for training (and particularly more low SNR events), resulting in a better model. Presumably, a more sophisticated trigger (e.g. cWB \rev{\cite{klimenko2008coherent}}) would improve the overall recall. One intriguing approach --- reserved for future work --- would be to use an unrelated deep learning method with a permissive threshold as the trigger. Note that this would be a starkly different regime for the trigger algorithm compared to current implementations: whereas most algorithms are designed for use in the high-precision regime, in the given example, the algorithm would aim to maximize recall while allowing a precision of less than 1\%.  

\section{Conclusion}

We conclude that CoAD has potential as a template-free search for GW events. The current demonstration uses a small pre-cleaned dataset; the next steps include adding phase information into the inputs, exploring more sophisticated triggering methods, \rev{studying dependence on event number and SNR, and ultimately deploying to a year-scale production dataset including transient noise.} Performance could also be improved through more powerful models (e.g. leveraging pre-trained foundation models), and additional frequency filtering with improved interpretability methods.

The parameters in this study reflect CBC search at the current Advanced LIGO. Two proposed projects --- the Einstein telescope \cite{maggiore2020science} and the Cosmic Explorer \cite{reitze2019cosmic} --- could increase the number of events (and thus also the signal rate) by a factor of 1000. Updates to Advanced LIGO could also increase the signal rate \cite{buchli2025improving}. In a high event-rate regime, CoAD has additional advantages: the larger number of events would make models easier to train, while simultaneously making it more difficult for alternative methods which require a background-only training set. A larger number of events would also unlock new possibilities in interpretability, including directly predicting physical templates. For example, a physics-informed neural network \cite{cai2021physics} could incorporate standard GW template matching procedures in the loss function to train the NN to output a physical template for each event.

Finally, we highlight that the concept presented here could find application to other fundamental physics searches where multiple detectors are involved, ranging from dark-matter detection \cite{amaral2025first} to multi-messenger astronomy, or even to multi-modal measurements within a single detector \cite{amram2021tag}. 

\begin{acknowledgments}
The author thanks Barry Barish, Phil Harris, and Jon Richardson for helpful comments, and is especially grateful to Katya Govorkova for providing SNR values for the Codabench data. This work is based on ideas supported by the U.S. Department of Energy, under DOE Contract No. DE-AC02-76SF00515 
\end{acknowledgments}

\appendix

\section{Data set}
\label{app:dataset}
All data comes from the Codabench challenge.  Datasets are split into training (80\%) and validation (20\%) partitions for both the signal and background. Testing consists of approximately 60k examples, with 9900 signal and 50k background, all held back from the training/validation partitions.  The high fraction of signal in the test set would overestimate precision, so we scale the observed false positive rate accordingly to estimate the precision for 0.2\% and 0.02\% test signal rates, respectively. 

For all datasets, we use a stochastic sampler that selects from the signal and background datasets at the desired rates.  As a result, with small batch sizes, some batches may consist of only background events.  We observe the stochastic gradients do not preclude training. For the 0.02\% event-rate training set, we would ideally have 500k background examples. Because we do not have sufficient backgrounds in the dataset, we instead use 50k examples for all cases, and resample the backgrounds at a correspondingly higher rate as needed. In all cases, we recycle the signal and background sets when they are exhausted (i.e. not necessarily at the same time) so we do not have epochs in the traditional sense.  Instead we train for a fixed number of batches (150k) with early stopping. We use the Adam optimizer, with dropout of 0.05, and a learning rate of $2.5 \times 10^{-4}$. For the 0.2\% event rate training we use batch-size of 512 examples, and for 0.02\% we use batch-size of 4096.  
We apply a regularization of $1 \times 10^{-2}$ to the logit output of the network (before application of a sigmoid in the loss function) to prevent runaway which could slow learning.

The data in the Codabench set is already cleaned \cite{campolongo2025building}. We further apply a short-time Fourier transform (STFT) with window size 128 and overlap of 96, and drop features beyond 1.6kHz, where the power goes to zero. We also tried an overlap of 64, which gave similar performance.

\section{Neural network architecture}
\label{app:nn_architecture}
The network is a small convolutional architecture shown in Table~\ref{tab:nn_params}. We use the same architecture and weights for both H and L networks. We have performed only rudimentary exploration of hyperparameters of the NN architecture, training parameters, and data pre-processing.

\begin{table}[h]
\centering
\begin{tabular}{l @{\hspace{1em}} c @{\hspace{1em}} r}
\hline
\textbf{Layer} & \textbf{Output Shape} & \textbf{Parameters} \\
\hline
\hline
Conv2d & [-1, 3, 52, 8] & 30 \\
Conv2d & [-1, 3, 52, 8] & 84 \\
Conv2d & [-1, 3, 26, 4] & 84 \\
Flatten & [-1, 312] & 0 \\
Dropout & [-1, 312] & 0 \\
Linear & [-1, 6] & 1,878 \\
Linear & [-1, 6] & 42 \\
Linear & [-1, 3] & 21 \\
\hline
 &  & Total: 2,139 \\
\hline
\end{tabular}
\caption{Architecture used for all neural networks in the study. `Conv2d' is a 2-dimensional convolution with a 3x3 kernel.}
\label{tab:nn_params}
\end{table}


\section{Loss function}
\label{app:loss}

The original CoAD formulation is given in Eq. 3 of \cite{humble2024coincident}. However, subsequent work \cite{oshea2024coincidence} found that the simpler correlation and covariance metrics --- while having fewer theoretical guarantees -- train more reliably. These two metrics can also be combined through an analogous $F_\beta$ construction to include combinations between the limits of covariance (identical to infinite $\beta$), and correlation (similar to low $\beta$). 
Note that while the optimal correlation score is always 1, the optimal covariance score decreases with lower anomaly rates, and in our case, the covariance scores are 2-3 orders of magnitude smaller than the correlation scores. The original work introduced the $\alpha$ parameter (an estimated anomaly rate) to account for this scaling.  In this work, we apply a simple scaling to each output, resulting in the final loss function: 
\begin{align}
\hat{F}_\beta&(s_H,s_L) = \nonumber\\ 
&\sum_i \frac{\alpha[i] (1+\beta[i]^2) \,\text{Corr}(s_H[i],s_L[i]) \text{Cov}(s_H[i],s_L[i])}{\beta[i]^2 \,\text{Corr}(s_H[i],s_L[i]) + \text{Cov}(s_H[i],s_L[i])}
\end{align}
where $s_H[i]$ and $s_L[i]$ are the sigmoids of the $i^{\text{th}}$ outputs of the two neural networks, and `Corr' and `Cov' are shorthands for correlation and covariance respectively. 

The simplest implementation of CoAD uses a single scalar prediction, but it is also possible to train simultaneously on multiple outputs corresponding to different $\beta$ values. The use of multiple outputs was originally introduced to enable fine-tuning of pre-trained networks, where it may be desirable for the network to retain knowledge of multiple signal types, e.g. if continually introducing seeds from the original training set. This is because with a single value the network predictions should collapse to one partition, which might select the seed signal rather than the true signal.  Multiple values should retain the seed information (low $\beta$) while also striving to include the new, rarer true signal (high $\beta$). However, even for direct training, the multiple outputs may encourage better structure in the networks latent space, but this requires further investigation.  The results presented in this work use $\beta = [0.01, 1, 100]$, and $\alpha=[1, 1, 1]$, which we found empirically to be most stable. Over-weighting the high-$\beta$ outputs using $\alpha = [1, 10, 10^2]$ generally resulted in worse results, perhaps because correlation training is more reliable. The use of higher $\alpha$ and $\beta$ values should be studied more in the future.


\section{Training and Model selection}
\label{app:model_selection}

\begin{figure}[htbp]
    \centering
    \vspace{-.1em}
    \includegraphics[width=1\linewidth]{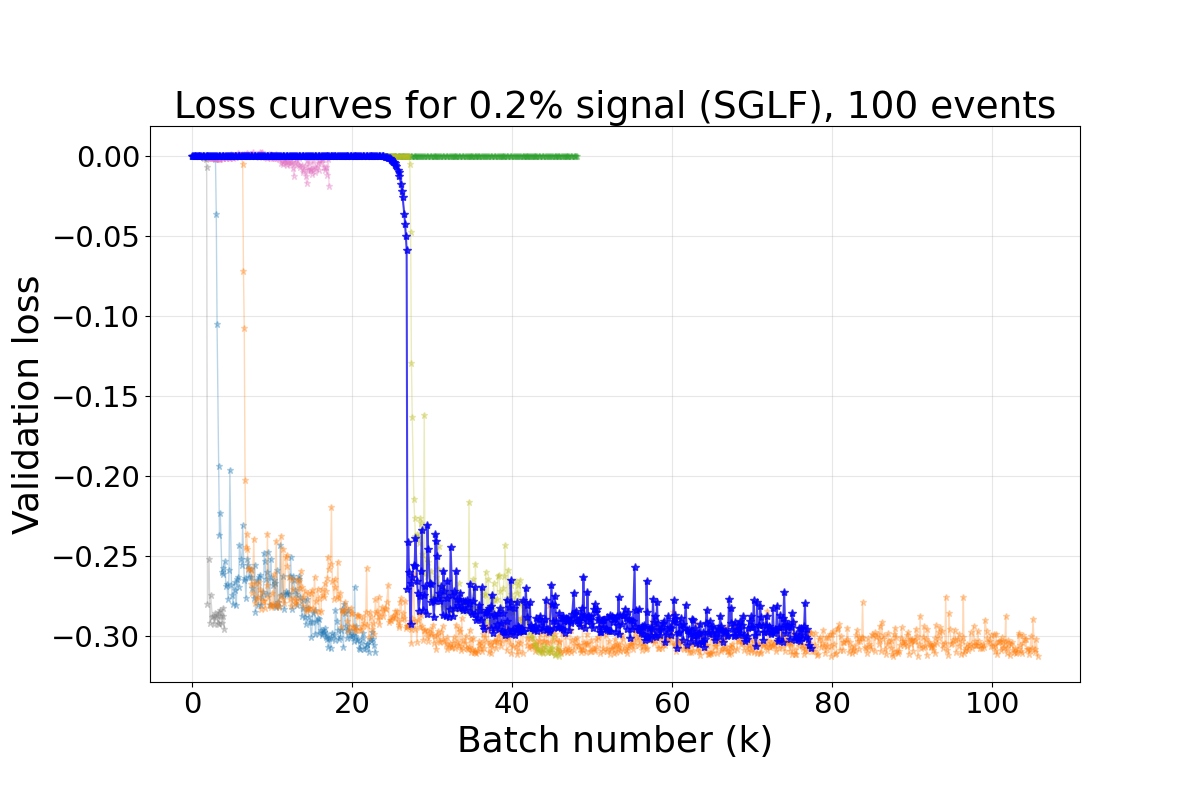}
    \vspace{-2em}
    \caption{Validation loss values as a function of batch for the 250 event, 0.2\% anomaly rate, SGLF model. The training curve for the final model is shown in blue.}
    \label{fig:loss_curves}
\end{figure}

CoAD loss functions have a tendency to become trapped in local minima. Because the network is small and training is relatively cheap, we take the approach of rerunning the training multiple times, and selecting the best model (e.g. see \cite{liang2025coincident}). The 0.2\% models were trained 10 times, and the 0.02\% model was trained 20 times.  The total training time for each case depends on the fraction of models that converge, but ranged from under 3 hours (0.2\%, 100 event case) to over 12 hours (0.02\% case) on a single T4 TPU. Loss curves for the 100 event, 0.2\%, SGLF model are given in Fig.~\ref{fig:loss_curves}.

\begin{figure}[htbp]
    \centering
    \vspace{-0.2em}
    \includegraphics[width=.95\linewidth]{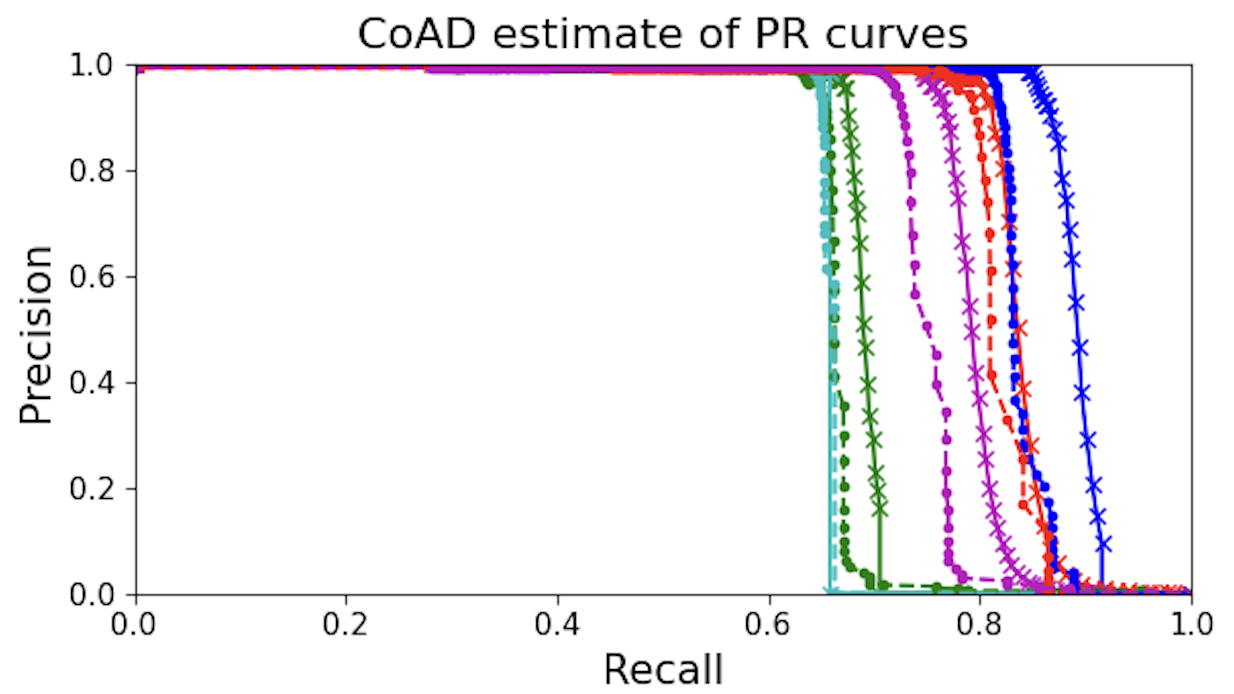}
    \vspace{-.2em}
    \caption{Comparison of the ground truth (crosses, solid line) and the CoAD unsupervised estimates (circles, dashed line) for the six best models trained in Fig.~\ref{fig:loss_curves}, \rev{with the color indicating the model}. The chosen model for Table~\ref{tab:results} is shown in blue.}
    \label{fig:pr_curve_coad_vs_gt}
\end{figure}

\begin{figure}[htbp]
    \centering
    \vspace{-0.1em}
    \includegraphics[width=.9\linewidth]{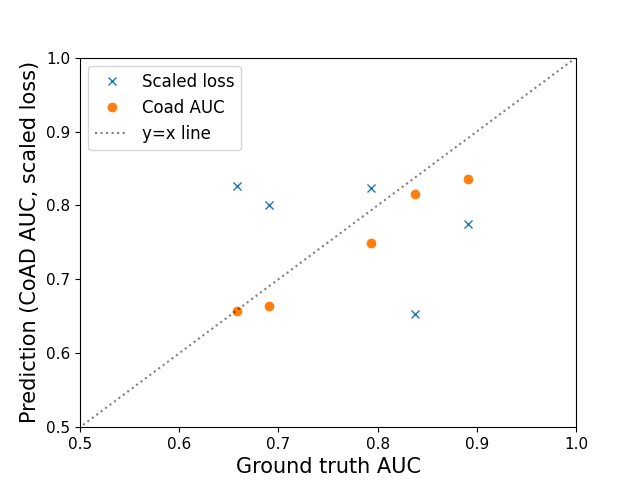}
    \caption{Comparison of two options for assessing model performance: the unsupervised CoAD AUC (orange circles) and a scaled version of the loss function (blue crosses). Compared to the loss values, the unsupervised CoAD AUC has significantly stronger correlation to the ground truth AUC. Points are shown for the 5 best models.  (The other 5 trained models failed to converge.) The CoAD estimate is also scaled to match the ground truth, since the true normalization is not known, but this does not affect the correlation.}
    \label{fig:coad_vs_gt_vs_loss}
\end{figure}

In the absence of labels, there is need for an alternative mechanism to select the best model produced from the multiple training runs. The simplest approach is to select the model with the lowest loss value. However, while the loss is designed to mimic the AUC metric, the two scores have differences.  Moreover, the loss is an average across the three $\beta$ values, and a model with good behavior on only one output would still be useful. Consequently, the model with the best loss does not always deliver the best performance. Instead, we can use the CoAD concept to formulate unsupervised versions of precision ($\hat{P}$) and recall ($\hat{R}$) as defined in \cite{humble2024coincident}.  From $\hat{P}$ and $\hat{R}$ we can then create an unsupervised analog of the AUC score by scanning thresholds. (Note that while $\hat{R}$ is unnormalized, this does not affect our ability to evaluate the relative performance of different models.) The unsupervised AUC metric is both computationally expensive and not directly differentiable, so it is not suitable for training the models. However, it is a more faithful approximation of our true metric, so we use the unsupervised AUC to both select the best model, and to select the best performing of the three outputs for each model.  (In principle we could combine the outputs, rather than selecting one, but have not explored that option.) Comparison of the unsupervised metrics and the ground truth AUC is shown in Figs.~\ref{fig:pr_curve_coad_vs_gt} and \ref{fig:coad_vs_gt_vs_loss}.

\section{Inference}
\label{app:inference}

At inference time, the predictions from the two networks are independent.  In principle, if only one datastream is available, the predictions from a single network can be used to identify GW events. In practice, combining the predictions from the two networks dramatically improves the precision. The two signals can be combined in various ways. We have considered two approaches: first, we can take the product of the two predictions, and then apply a threshold, i.e.: 

\begin{align} 
    \hat{y} &= 
    \begin{cases} 
    \text{True} & \text{if } s_H * s_L \geq \tau \\ 
    \text{False} & \text{otherwise} 
    \end{cases}
    \intertext{with threshold $\tau$. Alternatively, we can employ a logical AND operation where we select a threshold and require both predictions to be above that threshold:}
    \hat{y} &= 
    \begin{cases} 
    \text{True} & \text{if } s_H \geq \tau \text{ AND } s_L \geq \tau \\ 
    \text{False} & \text{otherwise} 
    \end{cases}
\end{align}
We observe that the product method has marginally better performance, but because the AND approach is more physical --- and makes it easier to develop a CoAD metric --- we have used that method throughout this work.  Note that in the AND approach, we could apply different thresholds $\tau_H$ and $\tau_L$ for the two networks, but for simplicity we have used a single threshold.



\bibliography{accel}

\end{document}